\documentclass[aps,prd,twocolumn,superscriptaddress]{revtex4}
\usepackage{epsfig,epsf}
\usepackage{amsmath}
\usepackage{amsthm}
\usepackage{amsfonts}
\usepackage{amssymb}
\usepackage{dsfont}
\usepackage{multirow}
\usepackage{appendix}
\usepackage{slashed}
\usepackage[active]{srcltx}
\usepackage{psfrag}

\begin{document}

\title{The doubly charmed pseudoscalar tetraquarks $T_{cc;\bar{s} \bar{s}%
}^{++}$ and $T_{cc;\bar{d} \bar{s}}^{++}$}
\date{\today}
\author{S.~S.~Agaev}
\affiliation{Institute for Physical Problems, Baku State University, Az--1148 Baku,
Azerbaijan}
\author{K.~Azizi}
\affiliation{Department of Physics, Do\v{g}u\c{s} University, Acibadem-Kadik\"{o}y, 34722
Istanbul, Turkey}
\author{B.~Barsbay}
\affiliation{Department of Physics, Kocaeli University, 41380 Izmit, Turkey}
\author{H.~Sundu}
\affiliation{Department of Physics, Kocaeli University, 41380 Izmit, Turkey}

\begin{abstract}
The mass and coupling of the doubly charmed $J^P=0^{-}$ diquark-antidiquark
states $T_{cc;\bar{s} \bar{s}}^{++}$ and $T_{cc;\bar{d} \bar{s}}^{++}$ that
bear two units of the electric charge are calculated by means of QCD
two-point sum rule method. Computations are carried out by taking into
account vacuum condensates up to and including terms of tenth dimension. The
dominant $S$-wave decays of these tetraquarks to a pair of conventional $%
D_{s}^{+}D_{s0}^{\ast +}(2317)$ and $D^{+}D_{s0}^{\ast +}(2317)$ mesons are
explored using QCD three-point sum rule approach, and their widths are
found. The obtained results $m_{T}=(4390~\pm 150)~\mathrm{MeV}$ and $\Gamma
=(302 \pm 113~\mathrm{MeV}$) for the mass and width of the state $T_{cc;\bar{%
s} \bar{s}}^{++}$, as well as spectroscopic parameters $\widetilde{m}%
_{T}=(4265\pm 140)~\mathrm{MeV}$ and $\widetilde{\Gamma }=(171~\pm 52)~%
\mathrm{MeV}$ of the tetraquark $T_{cc;\bar{d} \bar{s}}^{++}$ may be useful
in experimental studies of exotic resonances.
\end{abstract}

\maketitle

%\affiliation{School of Physics, Institute for Research in Fundamental Sciences (IPM),
%P.~O.~Box 19395-5531, Tehran, Iran}

%%%%%%%%%%%%%%%%%%%%%%%%%%%%%%%%%%%%%%%%%%%%%%%%%%%%%%%%%%%%%%%%%%%%%%%%%%%%%

\section{Introduction}

\label{sec:Introduction}
%%%%%%%%%%%%%%%%%%%%%%%%%%%%%%%%%%%%%%%%%%%%%%%%%%%%%%%%%%%%%%%%%%%%%%%%%%%%%%
The investigation of exotic mesons, i.e. particles either with unusual
quantum numbers that are not accessible in the quark-antiquark $\overline{q}%
q $ model or built of four valence quarks (tetraquarks) remains among
interesting and important topics in high energy physics. Existence of
multiquark hadrons does not contradict to first principles of QCD and was
theoretically predicted already by different authors \cite%
{Jaffe:1976ig,Weinstein:1982gc,Ader:1981db}. But only after experimental
discovery of the charmonium-like resonance $X(3872)$ by Belle Collaboration
in 2003 \cite{Choi:2003ue} the exotic hadrons became an object of rapidly
growing studies. In the years that followed, various collaborations reported
about observation of similar resonances in exclusive and inclusive hadronic
processes. Theoretical investigations also achieved remarkable successes in
interpretation of exotic hadrons by adapting existing methods to a new
situation and/or inventing new approaches for their studies. Valuable
experimental data collected during fifteen years passed from the discovery
of the $X(3872)$ resonance, as well as important theoretical works
constitute now the physics of the exotic hadrons \cite%
{Chen:2016qju,Chen:2016spr,Esposito:2016noz,Ali:2017jda,Olsen:2017bmm}.

One of the main problems in experimental investigations of the
charmonium-like resonances is separation of tetraquark's effects from
contributions of the conventional charmonium and its numerous excited
states. Indeed, it is natural to explain neutral resonances detected in an
invariant mass distribution of final mesons as ordinary charmonia: only
detailed analyses may reveal their exotic nature. But there are few classes
of tetraquarks which can not be confused with the charmonium states. The first
class of such particles are resonances that bear the electric charge: it is
evident that $\overline{q}q$ mesons are neutral particles. The first charged
tetraquarks $Z_{c}^{\pm }(4430)$ were observed in decays of the $B$ meson $%
B\rightarrow K\psi ^{\prime }\pi ^{\pm }$ as resonances in the $\psi
^{\prime }\pi ^{\pm }$ invariant mass distribution \cite{Choi:2007wga}.
Later other charged resonances such as $Z_{c}^{\pm }(3900)$ were discoreved,
as well.

The next group are resonances composed of more than two quark flavors. The
quark content of such states can be determined from analysis of their decay
products. The prominent member of this group is the resonance $X^{\pm
}(5568) $ which is presumably composed of four distinct quark flavors. It
was first observed in the $B_{s}^{0}\pi ^{\pm }$ invariant mass distribution
in the $B_{s}^{0}$ meson hadronic decay mode, and confirmed later with the $%
B_{s}^{0}$ meson's semileptonic decays by the D0 Collaboration \cite%
{D0:2016mwd,Abazov:2017poh}. However, the LHCb and CMS collaborations could
not provide an evidence for its existence from analysis of relevant
experimental data \cite{Aaij:2016iev,CMS:2016}. Therefore, the experimental
status of the $X(5568)$ resonance remains unclear and controversial.

Resonances carrying a double electric charge constitute another very
interesting class of exotic states, because the doubly charged resonances
cannot be explained as conventional mesons \cite%
{Ohkoda:2012hv,Esposito:2013fma}. The doubly charged particles may exist as
doubly charmed tetraquarks composed of the heavy diquark $cc$ and light
antidiquarks $\bar{s}\bar{s}$ or $\bar{d}\bar{s}$. In other words, they can
contain two or three quark flavors. The diquark $bb$ and antidiquark $%
\overline{u}\overline{u}$ can also bind to form the doubly charged resonance
$T_{bb;\bar{u}\bar{u}}^{--}$ containing only two quark spices. The states
built of four quarks of different flavors can carry a double charge, as well
\cite{Chen:2017rhl}. The doubly charged molecular compounds with the quark
content $\overline{Q}\overline{Q}qq$, where $Q$ is $c$ or $b$-quark were
analyzed in Ref. \cite{Ohkoda:2012hv}. In this paper the authors used the
heavy quark effective theory to derive interactions between heavy mesons and
coupled channel Schrodinger equations to find the bound and/or resonant
states with various quantum numbers. It was demonstrated that, \ for
example, $D$ and $D^{\ast }$ mesons can form doubly charged $P$-wave bound
state with $J^{P}=0^{-}$.

The class of exotic states composed of heavy $cc$ and $bb$ diquarks and
heavy or light antidiquarks attracted already interests of scientists. The
four-quark systems $QQ\bar{Q}\bar{Q}$ and $QQ\bar{q}\bar{q}$ were studied in
Ref.\ \cite{Ader:1981db,Lipkin:1986dw,Zouzou:1986qh} by adopting the
conventional potential model with additive pairwise interaction of
color-octet exchange type. The goal was to find four-quark states which are
stable against spontaneous dissociation into two mesons. It turned out that
within this approach there are not stable mesons built of only heavy quarks.
But the states $QQ\bar{q}\bar{q}$ may form the stable composites provided
the ratio $m_{Q}/m_{q}$ is large. The same conclusions were drawn from a
more general analysis in Ref.\ \cite{Carlson:1987hh}, where the only
assumption made about the confining potential was its finiteness when two
particles come close together. In accordance with predictions of this paper
the isoscalar $J^{P}=1^{+}$ tetraquark $T_{bb;\bar{u}\bar{d}}^{-}$ lies
below the two B-meson threshold and hence, can decay only weakly. The
situation with of $T_{cc;\bar{q}\bar{q}^{\prime }}$ and $T_{bc;\bar{q}\bar{q}%
^{\prime }}$ is not quite clear, but they may exist as unstable bound
states. The stability of the $QQ\overline{q}\overline{q}$ compounds in the
limit $m_{Q}\rightarrow \infty $ was studied in Ref.\ \cite{Manohar:1992nd}
as well.

Production mechanisms of the doubly charmed tetraquarks in the ion,
proton-proton and electron-positron collisions, as well as their possible
decay channels were also examined in the literature \cite%
{SchaffnerBielich:1998ci,DelFabbro:2004ta,Lee:2007tn,Hyodo:2012pm}. The
chiral quark models, the dynamical and relativistic quark models were
employed to investigate properties (mainly to compute masses) of these
exotic mesons \cite{Pepin:1996id,Cui:2006mp,Vijande:2006jf,Ebert:2007rn}.
The similar problems were addressed in the context of QCD sum rule method as
well. The masses of the axial-vector states $T_{QQ;\bar{u}\bar{d}}$ were
extracted from the two-point sum rules in Ref.\ \cite{Navarra:2007yw}. The
mass of the tetraquark $T_{bb;\bar{u}\bar{d}}^{-}$ in accordance with this
work amounts to $10.2\pm 0.3\ \mathrm{GeV}$, and is below the open bottom
threshold. Within the same framework masses of the $QQ\bar{q}\bar{q}$ states
with the spin-parity $0^{-},\ 0^{+},\ 1^{-}$ and $1^{+}$ were computed in
Ref.\ \cite{Du:2012wp}.

Recently interest to double-charm and double-bottom tetraquarks renewed
after discovery of the doubly charmed baryon $\Xi _{cc}^{++}=ccu$ by the
LHCb Collaboration \cite{Aaij:2017ueg}. Thus, in Ref.\ \cite%
{Karliner:2017qjm} the masses of the tetraquarks $T_{bb;\overline{u}%
\overline{d}}^{-}$ and $T_{cc;\overline{u}\overline{d}}^{+}$ were estimated
in the context of a phenomenological model. The obtained prediction for $%
m=10389\pm 12\ \mathrm{MeV}$ confirms that the isoscalar state $T_{bb%
\overline{u}\overline{d}}^{-}$ with spin-parity $J^{P}=1^{+}$ is stable
against strong and electromagnetic decays, whereas the tetraquark $T_{cc%
\overline{u}\overline{d}}^{+}$ lies above the open charm threshold $%
D^{0}D^{\ast +}$ and can decay to these mesons. The various aspects of
double- and fully-heavy tetraquarks were also considered in Refs.\ \cite%
{Luo:2017eub,Eichten:2017ffp,Wang:2017dtg,Ali:2018ifm,Karliner:2016zzc,Bai:2016int, Richard:2017vry,Wu:2016vtq,Wang:2017jtz,Anwar:2017toa,Esposito:2018cwh,Yan:2018gik,Xing:2018bqt}%
. Works devoted to investigation of the hidden-charm (-bottom) tetraquarks
containing $c\overline{c}$ ($b\overline{b}$) may also provide interesting
information on properties of the heavy exotic states (see Ref.\ \cite%
{Chen:2016oma} and references therein).

The masses of the doubly charged exotic mesons built of four different quark
flavors were extracted from QCD sum rules in Ref.\ \cite{Chen:2017rhl}. The
spectroscopic parameters and full width of the scalar, pseudoscalar and
axial-vector doubly charged charm-strange tetraquarks $Z_{\bar{c}s}=[sd][%
\bar{u}\overline{c}]$ were calculated in Ref.\ \cite{Agaev:2017oay}. It was
shown that width of these compounds evaluated using their strong decay
channels ranges from $\Gamma _{\mathrm{PS}}=38.10\ \mathrm{MeV}$ in the case
of the pseudoscalar resonance till $\Gamma _{\mathrm{S}}=66.89\ \mathrm{MeV}$
for the scalar state, which is typical for most of the diquark-antidiquark
resonances.

In the present work we explore the pseudoscalar tetraquarks $T_{cc;\bar{s}%
\bar{s}}^{++}$ and $T_{cc;\bar{d}\bar{s}}^{++}$ that are doubly charmed and,
at the same time doubly charged exotic mesons. Their masses and couplings
are calculated using QCD two-point sum rules approach which is the powerful
quantitative method to analyze properties of hadrons including exotic states
\cite{Shifman:1978bx,Shifman:1978by}. Since the tetraquarks under discussion
are not stable and can decay strongly in $S$-wave to $D_{s}^{+}D_{s0}^{\ast
+}(2317)$ and $D^{+}D_{s0}^{\ast +}(2317)$ mesons we calculate also widths
of these channels. To this end, we utilize QCD three-point sum rule method
to compute the strong couplings $G_{s}$ and $G_{d}$ corresponding to the
vertices $T_{cc;\overline{s}\overline{s}}^{++}D_{s}^{+}D_{s0}^{\ast +}(2317)$
and $T_{cc;\overline{d}\overline{s}}^{++}D^{+}D_{s0}^{\ast +}(2317)$,
respectively. Obtained information on $G_{s}$ and $G_{d}$, as well as
spectroscopic parameters of the tetraquarks are applied as key ingredients
to evaluate the partial decay widths $\Gamma \lbrack T_{cc;\overline{s}%
\overline{s}}^{++}\rightarrow D_{s}^{+}D_{s0}^{\ast +}(2317)]$ and $%
\widetilde{\Gamma }[T_{cc;\overline{d}\overline{s}}^{++}\rightarrow
D^{+}D_{s0}^{\ast +}(2317)]$.

This work is organized in the following way: In the section \ref{sec:Mass}
we calculate the masses and couplings of the pseudoscalar tetraquarks using
the two-point sum rule method by including into analysis the quark, gluon
and mixed condensates up to dimension ten. The spectroscopic parameters of
these resonances are employed in Sec.\ \ref{sec:Decays} to evaluate strong
couplings and widths of the $T_{cc;\overline{s}\overline{s}}^{++}$ and $%
T_{cc;\overline{d}\overline{s}}^{++}$ states' $S$-wave strong decays. The
section \ref{sec:Conc} is reserved for analysis and our concluding remarks.
The Appendix contains explicit expressions of the correlation functions used
in calculations of the spectroscopic parameters and strong coupling of the
tetraquark $T_{cc;\overline{s}\overline{s}}^{++}$.

%%%%%%%%%%%%%%%%%%%%%%%%%%%%%%%%%%%%%%%%%%%%%%%%%%%%%%%%%%%%%%%%%%%%%%%%%%%%5

\section{The spectroscopy of the $J^P=0^{-}$ tetraquarks $T_{cc;\overline{s}%
\overline{s}}^{++}$ and $T_{cc;\overline{d}\overline{s}}^{++}$}

\label{sec:Mass}
%%%%%%%%%%%%%%%%%%%%%%%%%%%%%%%%%%%%%%%%%%%%%%%%%%%%%%%%%%%%%%%%%%%%%%%%%%%%%%%%%%%
One of the effective tools to evaluate the masses and couplings of the
tetraquarks $T_{cc;\overline{s}\overline{s}}^{++}$ and $T_{cc;\overline{d}%
\overline{s}}^{++}$ is QCD two-point sum rule method. In this section we
present in a detailed form calculation of these parameters in the case of
the diquark-antidiquark $T_{cc;\overline{s}\overline{s}}^{++}$ and provide
only final results for the second state $T_{cc;\overline{d}\overline{s}%
}^{++} $.

The basic quantity in the sum rule calculations is the correlation function
chosen in accordance with a problem under consideration. The best way to
derive the sum rules for the mass and coupling is analysis of the two-point
correlation function
\begin{equation}
\Pi (p)=i\int d^{4}xe^{ipx}\langle 0|\mathcal{T}\{J(x)J^{\dagger
}(0)\}|0\rangle ,  \label{eq:CF1}
\end{equation}%
where $J(x)$ in the interpolating current for the isoscalar $J^{P}=0^{-}$
state $T_{cc;\overline{s}\overline{s}}^{++}$. It can be defined in the
following form \cite{Du:2012wp}
\begin{eqnarray}
&&J(x)=c_{a}^{T}(x)Cc_{b}(x)\left[ \overline{s}_{a}(x)\gamma _{5}C\overline{s%
}_{b}^{T}(x)+\overline{s}_{b}(x)\gamma _{5}C\overline{s}_{a}^{T}(x)\right] ,
\notag \\
&&  \label{eq:Curr1}
\end{eqnarray}%
where $C$ is the charge conjugation operator, $a$ and $b$ are color indices.
The interpolating current for the isospinor tetraquark $T_{cc;\overline{d}%
\overline{s}}^{++}$ is given by the similar expression
\begin{eqnarray}
&&\widetilde{J}(x)=c_{a}^{T}(x)Cc_{b}(x)\left[ \overline{d}_{a}(x)\gamma
_{5}C\overline{s}_{b}^{T}(x)+\overline{d}_{b}(x)\gamma _{5}C\overline{s}%
_{a}^{T}(x)\right] .  \notag \\
&&  \label{eq:Curr2}
\end{eqnarray}%
The currents $J(x)$ and $\widetilde{J}(x)$ have symmetric color structure $[%
\mathbf{6}_{\mathbf{c}}]_{cc}\otimes \lbrack \overline{\mathbf{6}}_{\mathbf{c%
}}]_{\overline{s}\overline{s}}$ and are composed of the heavy pseudoscalar
diquark and light scalar antidiquark. There are other interpolating currents
with $J^{P}=0^{-}$ but composed, for example, of the heavy scalar diquark
and light pseudoscalar antidiquark \cite{Du:2012wp}. To describe the
tetraquarks $T_{cc;\overline{s}\overline{s}}^{++}$ and $T_{cc;\overline{d}%
\overline{s}}^{++}$ one can also use linear combinations of these currents.
In general, different currents may modify the results for the spectroscopic
parameters of the tetraquarks under consideration. In the present work we
restrict our analysis by the interpolating currents $J(x)$ and $\widetilde{J}%
(x)$ bearing in mind that among various diquarks the scalar ones are most
tightly bound states.

The QCD sum rule method implies calculation of the correlation function $\Pi
(p)$ using the phenomenological parameters of the tetraquark $T_{cc;%
\overline{s}\overline{s}}^{++}$, i. e. its mass $m_{T}$ and coupling $f_{T}$
from one side, and computation of $\Pi (p)$ in terms of the quark
propagators from another side. Equating expressions obtained by this way and
invoking the quark-hadron duality it is possible to derive the sum rules to
evaluate $m_{T}$ and $f_{T}$.

We assume that the phenomenological side of the sum rules can be
approximated by a single pole term. In the case of the multiquark systems
this approach has to be used with some caution, because the physical side
receives contribution also from two-hadron reducible terms. In fact, the
relevant interpolating current couples not only to the tetraquark
(pentaquark), but also to the two-hadron continuum lying below the mass of
the multiquark system \cite{Kondo:2004cr,Lee:2004xk}. These terms can be
either subtracted from the sum rules or included into parameters of the pole
term. The first method was employed mainly in investigating the pentaquarks
\cite{Lee:2004xk,Sarac:2005fn}, whereas the second approach was used to
study the tetraquarks \cite{Wang:2015nwa}. It turns out that the
contribution of the two-meson continuum generates the finite width $\Gamma
(p^{2})$ of the tetraquark and leads to the modification
\begin{equation}
\frac{1}{m_{T}^{2}-p^{2}}\rightarrow \frac{1}{m_{T}^{2}-p^{2}-i\sqrt{p^{2}}%
\Gamma (p^{2})}.  \label{eq:Mod}
\end{equation}%
These effects, properly taken into account in the sum rules, rescale the
coupling $f_{T}$ and leave untouched the mass of the tetraquark $m_{T}$.

In all cases explored in Refs. \cite{Lee:2004xk,Sarac:2005fn,Wang:2015nwa}
the two-hadron continuum effects were found small and negligible. Therefore,
to derive the phenomenological side of the sum rules we use the zero-width
single-pole approximation and demonstrate in the section \ref{sec:Conc} the
self-consistency of the obtained results by explicit computations.

In the context of this approach the correlation function $\Pi ^{\mathrm{Phys}%
}(p)$ takes a simple form%
\begin{equation}
\Pi ^{\mathrm{Phys}}(p)=\frac{\langle 0|J|T(p)\rangle \langle
T(p)|J^{\dagger }|0\rangle }{m_{T}^{2}-p^{2}}+\ldots ,  \label{eq:Phys1}
\end{equation}%
where by dots we indicate contribution of higher resonances and continuum
states. This formula can be simplified further by introducing the matrix
element%
\begin{equation}
\langle 0|J|T(p)\rangle =\frac{m_{T}^{2}f_{T}}{\mathcal{M}},  \label{eq:Mel1}
\end{equation}%
where $\mathcal{M=}2(m_{c}+m_{s})$. After some simple manipulations we get%
\begin{equation}
\Pi ^{\mathrm{Phys}}(p)=\frac{m_{T}^{4}f_{T}^{2}}{\mathcal{M}^{2}}\frac{1}{%
m_{T}^{2}-p^{2}}+\ldots .  \label{eq:Phys2}
\end{equation}%
It is seen that the Lorentz structure of the correlation function is trivial
and there is only a term proportional to $I$ . The invariant amplitude $\Pi
^{\mathrm{Phys}}(p^{2})=$ $m_{T}^{4}f_{T}^{2}/[\mathcal{M}%
^{2}(m_{T}^{2}-p^{2})]$ corresponding to this structure constitutes the
physical side of the sum rule. In order to suppress effects coming from
higher resonances and continuum states one has to apply to $\Pi ^{\mathrm{%
Phys}}(p^{2})$ the Borel transformation which leads to
\begin{equation*}
\mathcal{B}\Pi ^{\mathrm{Phys}}(p^{2})\equiv \Pi ^{\mathrm{Phys}}(M^{2})=%
\frac{m_{T}^{4}f_{T}^{2}e^{-m_{T}^{2}/M^{2}}}{\mathcal{M}^{2}}
\end{equation*}%
with $M^{2}$ being the Borel parameter.

The second side of the required equality $\Pi ^{\mathrm{OPE}}(p)$ is
accessible through computation of Eq.\ (\ref{eq:CF1}) using the explicit
expression of the interpolating current (\ref{eq:Curr1}) and contracting
quark fields under the time ordering operator $\mathcal{T}$. The expression
of $\Pi ^{\mathrm{OPE}}(p)$ in terms of quarks' propagators is written down
in the Appendix \ref{sec:App}. We employ the heavy $c$ and light $s$-quark
propagators, explicit expressions of which can be found in Ref.\ \cite%
{Agaev:2017tzv}, for example. The calculations are carried out at the
leading order of the perturbative QCD by taking into account quark, gluon
and mixed condensates up to dimension ten.

The invariant amplitude $\Pi ^{\mathrm{OPE}}(p^{2})$ can be written down in
terms of the spectral density $\rho (s)$%
\begin{equation}
\Pi ^{\mathrm{OPE}}(p^{2})=\int_{{\mathcal{M}^{2}}}^{\infty }\frac{\rho (s)}{%
s-p^{2}}ds.  \label{eq:OPE2}
\end{equation}%
After equating $\Pi ^{\mathrm{Phys}}(M^{2})$ to the Borel transform of $\Pi
^{\mathrm{OPE}}(p^{2})$ and performing the continuum subtraction we get a
first expression that can be used to derive the sum rules for the mass and
coupling. The second equality can be obtained from the first one by applying
the operator $d/d(-1/M^{2})$. Then it is not difficult we find the sum rules
for $m_{T}$ and $f_{T}$
\begin{equation}
m_{T}^{2}=\frac{\int_{\mathcal{M}^{2}}^{s_{0}}ds\rho (s)se^{-s/M^{2}}}{\int_{%
\mathcal{M}^{2}}^{s_{0}}ds\rho (s)e^{-s/M^{2}}},  \label{eq:SR1}
\end{equation}%
and
\begin{equation}
f_{T}^{2}=\frac{\mathcal{M}^{2}}{m_{T}^{4}}\int_{\mathcal{M}%
^{2}}^{s_{0}}ds\rho (s)e^{(m_{T}^{2}-s)/M^{2}}.  \label{eq:SR2}
\end{equation}%
In Eqs.\ (\ref{eq:SR1}) and (\ref{eq:SR2}) $s_{0}$ is the continuum
threshold parameter introduced during the subtraction procedure: it
separates the ground-state and continuum contributions.

The sum rules for the mass and coupling depend on numerous parameters, which
should be fixed to carry out numerical analysis. Below we write down the
quark, gluon and mixed condensates
\begin{eqnarray}
&&\langle \bar{q}q\rangle =-(0.24\pm 0.01)^{3}\ \mathrm{GeV}^{3},\ \langle
\bar{s}s\rangle =0.8\ \langle \bar{q}q\rangle ,  \notag \\
&&m_{0}^{2}=(0.8\pm 0.1)\ \mathrm{GeV}^{2},\ \langle \overline{q}g_{s}\sigma
Gq\rangle =m_{0}^{2}\langle \overline{q}q\rangle ,  \notag \\
&&\langle \overline{s}g_{s}\sigma Gs\rangle =m_{0}^{2}\langle \bar{s}%
s\rangle ,  \notag \\
&&\langle \alpha _{s}G^{2}\rangle =(6.35\pm 0.35)\cdot 10^{-2}\,\mathrm{GeV}%
^{4},  \notag \\
&&\langle g_{s}^{3}G^{3}\rangle =(0.57\pm 0.29)\ \mathrm{GeV}^{6},
\label{eq:Param}
\end{eqnarray}%
used in numerical computations. For the gluon condensate $\langle \alpha
_{s}G^{2}\rangle $ we employ its new average value presented recently in
Ref.\ \cite{Narison:2018dcr}, whereas $\langle g_{s}^{3}G^{3}\rangle $ is
borrowed from Ref.\ \cite{Narison:2015nxh}. For the masses of the $c$ and $s$%
-quarks
\begin{equation}
\mathrm{\ }m_{c}=1.275_{-0.035}^{+0.025}~\mathrm{GeV,\ }m_{s}=95_{-3}^{+9}~%
\mathrm{MeV,}  \label{eq:QuarkM}
\end{equation}%
we utilize the information from Ref.\ \cite{Tanabashi:2018}.

Besides, the sum rules contain also the auxiliary parameters $M^{2}$ and $%
s_{0}$ which may be varied inside of some regions and must satisfy standard
restrictions of the sum rules computations. The analysis demonstrates that
the working windows
\begin{equation}
M^{2}=(4.7,\ 7.0)\ \mathrm{GeV}^{2},\ s_{0}=(22,\ 24)\ \mathrm{GeV}^{2},
\label{eq:Reg1}
\end{equation}%
meet constraints imposed on $M^{2}$ and $s_{0}$. Indeed, the pole
contribution (\textrm{PC}) changes within limits $55\%-22\%$ when one varies
$M^{2}$ from its minimal to maximal allowed values: the higher limit of the
Borel parameter is fixed namely from exploration of the pole contribution.
The lower bound for $M^{2}$ stems from the convergence of the operator
product expansion (OPE)
\begin{equation}
R(M^{2})=\frac{\Pi ^{\mathrm{Dim(8+9+10)}}(M^{2},\ s_{0})}{\Pi (M^{2},\
s_{0})}<0.05,  \label{eq:Conv}
\end{equation}%
where $\Pi (M^{2},\ s_{0})$ is the subtracted Borel transform of $\Pi ^{%
\mathrm{OPE}}(p^{2})$, and $\Pi ^{\mathrm{Dim(8+9+10)}}(M^{2},\ s_{0})$ is
contribution of the last three terms in expansion of the correlation
function. At minimal $M^{2}$ the ratio $R$ is equal to $R(4.7~\mathrm{GeV}%
^{2})=0.018$ which proves the nice convergence of the sum rules. Moreover,
at $M^{2}=4.7~\mathrm{GeV}^{2}$ the perturbative contribution amounts to
more than $88\%$ of the full result and considerably exceeds the
nonperturbative contributions.

The mass $m_{T}$ and coupling $f_{T}$ extracted from the sum rules should be
stable under variation of the parameters $M^{2}$ and $s_{0}$. However in
calculations these quantities show a sensitivity to the choice both of $%
M^{2} $ and $s_{0}$. Therefore, when choosing the intervals for $M^{2}$ and $%
s_{0}$ we demand maximal stability of $m_{T}$ and $f_{T}$ on these
parameters. As usual, the mass $m_{T}$ of the tetraquark is more stable
against variation of $M^{2}$ and $s_{0}$ which is seen from Figs.\ \ref%
{fig:Mass} and \ref{fig:Coupling}. This fact has simple explanation: the sum
rule for the mass is given by Eq.\ (\ref{eq:SR1}) as the ratio of two
integrals, therefore their uncertainties partly cancel each other smoothing
dependence of $m_{T}$ on the Borel and continuum threshold parameters. The
coupling $f_{T}$ is more sensitive to the choice of $M^{2}$ and $s_{0}$,
nevertheless corresponding ambiguities do not exceed $20\%$ staying within
limits typical for sum rules calculations.

\begin{widetext}

\begin{figure}[h!]
\begin{center} \includegraphics[%
totalheight=6cm,width=8cm]{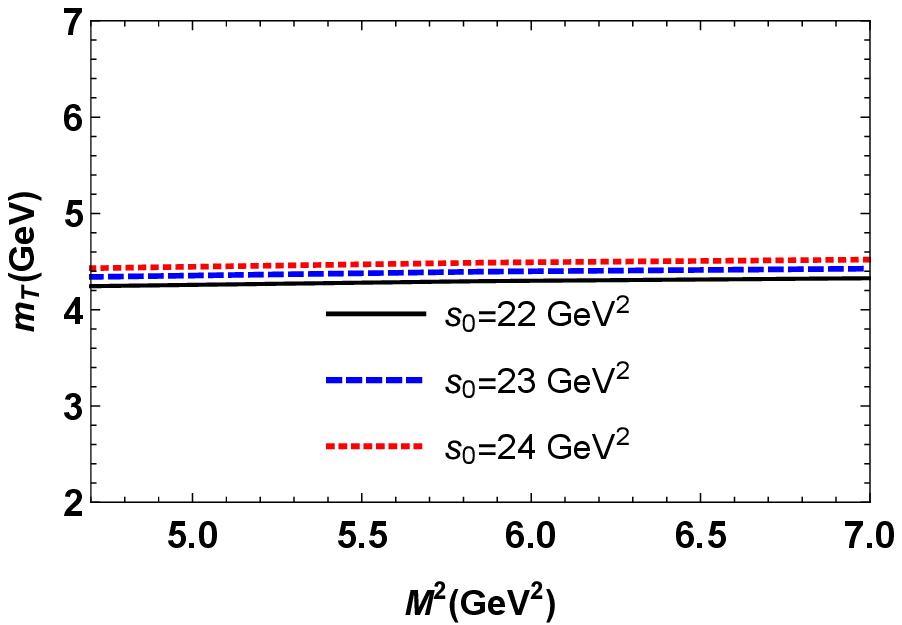}\,\,
\includegraphics[
totalheight=6cm,width=8cm]{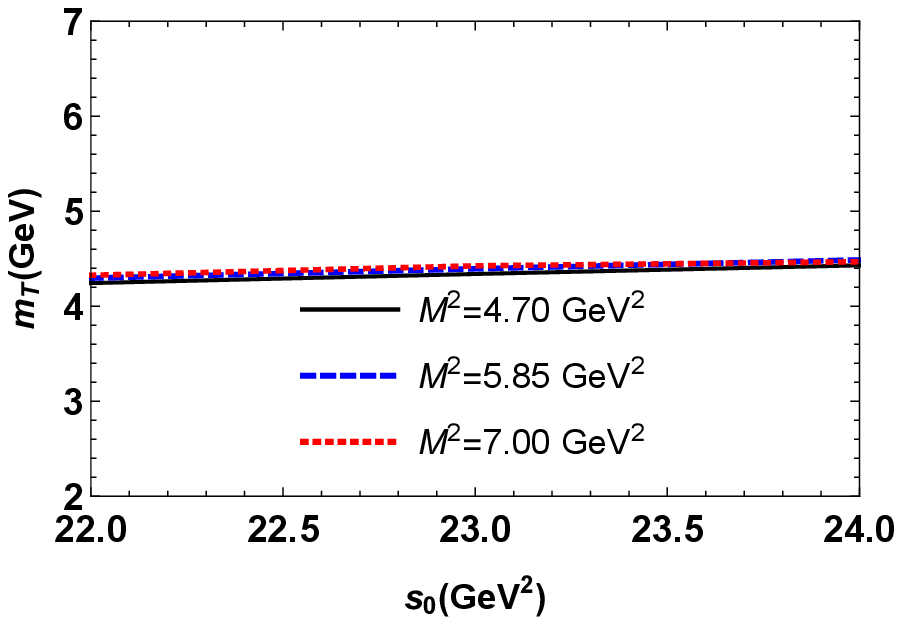}
\end{center}
\caption{ The mass of the tetraquark $T_{cc;\overline{s}%
\overline{s}}^{++}$ as a function of  the Borel parameter (left), and continuum threshold
parameter (right).}
\label{fig:Mass}
\end{figure}
\begin{figure}[h!]
\begin{center}
\includegraphics[%
totalheight=6cm,width=8cm]{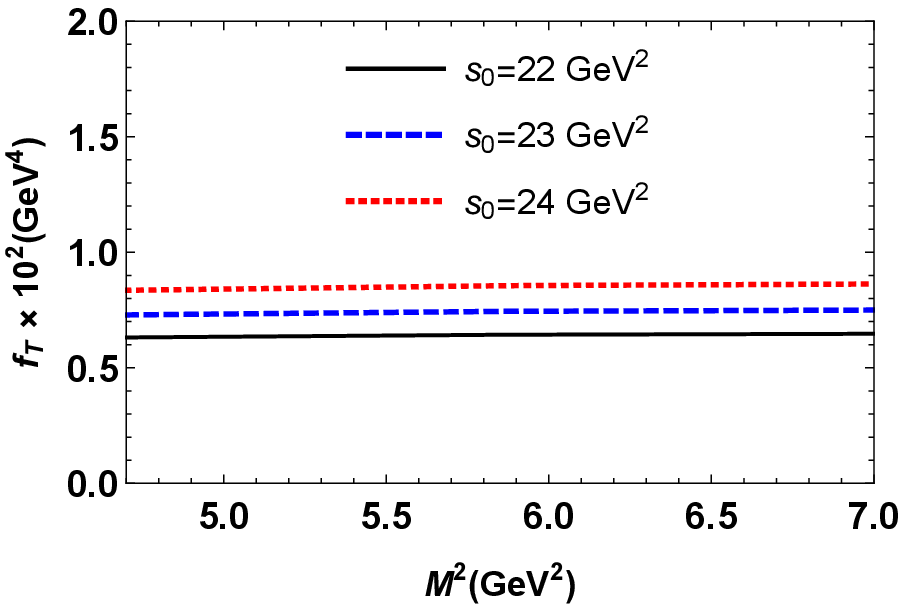}\,\,
\includegraphics[
totalheight=6cm,width=8cm]{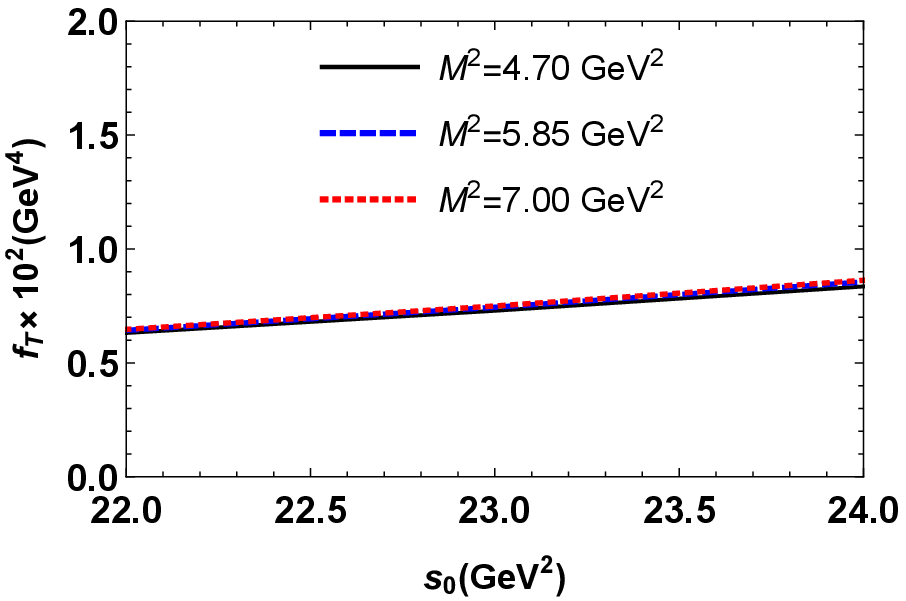}
\end{center}
\caption{ The dependence of the coupling $f_{T}$ on  the Borel (left), and continuum threshold
(right) parameters.}
\label{fig:Coupling}
\end{figure}

\end{widetext}

From performed analysis for the mass and coupling of the tetraquark $T_{cc;%
\overline{s}\overline{s}}^{++}$ we find
\begin{eqnarray}
m_{T} &=&(4390~\pm 150)~\mathrm{MeV},  \notag \\
f_{T} &=&(0.74\pm 0.14)\cdot 10^{-2}\ \mathrm{GeV}^{4}.  \label{eq:CMass1}
\end{eqnarray}%
The similar investigations of the $T_{cc;\overline{d}\overline{s}}^{++}$\
lead to predictions
\begin{eqnarray}
\widetilde{m}_{T} &=&(4265\pm 140~)~\mathrm{MeV},  \notag \\
\widetilde{f}_{T} &=&(0.62\pm 0.10)\cdot 10^{-2}~\mathrm{GeV}^{4},
\label{eq:CMass2}
\end{eqnarray}%
which have been obtained using the working regions
\begin{equation}
M^{2}=(4.5,\ 6.5)\ \mathrm{GeV}^{2},\ s_{0}=(21,\ 23)\ \mathrm{GeV}^{2}.
\label{eq:Reg2}
\end{equation}%
Let us note that in calculations of $\widetilde{m}_{T}$ and $\widetilde{f}%
_{T}$ the pole contribution $\mathrm{PC}\ $ changes within limits $59\%-27\%$%
. Contribution of the last three terms to the corresponding correlation
function at the point $M^{2}=4.5\ \mathrm{GeV}^{2}$\ amounts to $1.8\%$ of
the total result, which demonstrates convergence of the sum rules.

The spectroscopic parameters of the tetraquarks $T_{cc;\overline{s}\overline{%
s}}^{++}$ and $T_{cc;\overline{d}\overline{s}}^{++}$ obtained here will be
utilized in the next section to determine width of their decay channels.

%%%%%%%%%%%%%%%%%%%%%%%%%%%%%%%%%%%%%%%%%%%%%%%%%%%%%%%%%%%%%%%%%%%%%%%%%%%%5

\section{The decays $T_{cc;\overline{s}\overline{s}}^{++}\rightarrow
D_{s}^{+}D_{s0}^{\ast +}(2317)$ and $T_{cc;\overline{d}\overline{s}%
}^{++}\rightarrow D^{+}D_{s0}^{\ast +}(2317)$}

\label{sec:Decays}
%%%%%%%%%%%%%%%%%%%%%%%%%%%%%%%%%%%%%%%%%%%%%%%%%%%%%%%%%%%%%%%%%%%%%%%%%%%%%%%%%%%%%%%%%%%%%%%%%
The masses of the tetraquarks $T_{cc;\overline{s}\overline{s}}^{++}$ and $%
T_{cc;\overline{d}\overline{s}}^{++}$ allow us to fix their possible decay
channels. Thus, the tetraquark $T_{cc;\overline{s}\overline{s}}^{++}$ in $S$%
-wave decays to a pair of conventional mesons $D_{s}^{+}$ and $D_{s0}^{\ast
+}(2317)$, whereas the process $T_{cc;\overline{d}\overline{s}%
}^{++}\rightarrow D^{+}D_{s0}^{\ast +}(2317)$ is the main $S$-wave decay
channel of $T_{cc;\overline{d}\overline{s}}^{++}$. In fact, the threshold
for production of these particles can be easily calculated employing their
masses (see, Table \ref{tab:Param}): for production of the mesons $%
D_{s}^{+}D_{s0}^{\ast +}(2317)$ it equals to $(4286.04\pm 0.60)~\mathrm{%
MeV,}$ and for $D^{+}D_{s0}^{\ast +}(2317)$ amounts to $(4187.35\pm 0.60)~\mathrm{MeV}$. 
We see that the masses of the tetraquarks $T_{cc;\overline{s}%
\overline{s}}^{++}$ and $T_{cc;\overline{d}\overline{s}}^{++}$ are
approximately $104\ \mathrm{MeV}$ and $78\ \mathrm{MeV}$ above these
thresholds. There are also kinematically allowed $P$-wave decay modes of the
tetraquarks $T_{cc;\overline{s}\overline{s}}^{++}$ and $T_{cc;\overline{d}%
\overline{s}}^{++}$. Thus, the tetraquark $T_{cc;\overline{s}\overline{s}%
}^{++}$ through $P$-wave can decay to the final state $D_{s}^{+}D_{s}^{\ast
+}$, whereas for $T_{cc;\overline{d}\overline{s}}^{++}$ these channels are $%
D^{+}D_{s}^{\ast +}$ and $D^{\ast +}D_{s}^{+}$.  In the present work we
limit ourselves by considering only the $S$-wave decays of these tetraquarks.

In the present section we calculate the strong coupling form factor $G_{s}$
of the vertex $T_{cc;\overline{s}\overline{s}}^{++}\rightarrow
D_{s}^{+}D_{s0}^{\ast +}(2317)$ and find the width of the corresponding
decay channel $\Gamma \lbrack T_{cc;\overline{s}\overline{s}%
}^{++}\rightarrow D_{s}^{+}D_{s0}^{\ast +}(2317)]$. We provide also our
final predictions for $G_{d}$ and $\widetilde{\Gamma }[T_{cc;\overline{d}%
\overline{s}}^{++}\rightarrow D^{+}D_{s0}^{\ast +}(2317)]$ omitting details
of calculations which can easily be reconstructed from analysis of the first
process.

We use the three-point correlation function%
\begin{eqnarray}
\Pi (p,p^{\prime }) &=&i^{2}\int d^{4}xd^{4}ye^{i(p^{\prime }y-px)}\langle 0|%
\mathcal{T}\{J^{D_{s0}}(y)  \notag \\
&&\times J^{D_{s}}(0)J^{\dagger }(x)\}|0\rangle ,  \label{eq:CF2}
\end{eqnarray}%
to find the sum rule and extract the strong coupling $G_{s}$. Here $%
J^{D_{s}}(x)$ and $J^{D_{s0}}(x)$ are the interpolating currents for the
mesons $D_{s}^{+}$ and $D_{s0}^{\ast +}(2317)$, respectively. The
four-momenta of the tetraquark $T_{cc;\overline{s}\overline{s}}^{++}$ and
meson $D_{s0}^{\ast +}(2317)$ are $p$ and $p^{\prime }$: the momentum of the
meson $D_{s}^{+}$ then equals to $q=p-p^{\prime }$.

We define the interpolating currents of the mesons $D_{s}^{+}$ and $%
D_{s0}^{\ast +}(2317)$ in the following way%
\begin{equation}
J^{D_{s}}(x)=\overline{s}^{i}(x)i\gamma _{5}c^{i}(x),\ \ J^{D_{s0}}(x)=%
\overline{s}^{j}(x)c^{j}(x).  \label{eq:Curr3}
\end{equation}%
By isolating the ground-state contribution to the correlation function, for $%
\Pi ^{\mathrm{Phys}}(p,p^{\prime })$ we get%
\begin{eqnarray}
&&\Pi ^{\mathrm{Phys}}(p,p^{\prime })=\frac{\langle
0|J^{D_{s0}}|D_{s0}^{\ast }(p^{\prime })\rangle \langle
0|J^{D_{s}}|D_{s}(q)\rangle }{(p^{\prime
2}-m_{D_{s0}}^{2})(q^{2}-m_{D_{s}}^{2})}  \notag \\
&&\times \frac{\langle D_{s}(q)D_{s0}^{\ast }(p^{\prime })|T(p)\rangle
\langle T(p)|J^{^{\dagger }}|0\rangle }{(p^{2}-m_{T}^{2})}+\ldots ,
\label{eq:CF4}
\end{eqnarray}%
where the dots again stand for contributions of higher excited states and
continuum.

The correlation function $\Pi ^{\mathrm{Phys}}(p,p^{\prime })$ can be
further simplified by expressing matrix elements in terms of the mesons'
physical parameters. To this end we introduce the matrix elements
\begin{eqnarray}
\langle 0|J^{D_{s}}|D_{s}\rangle &=&\frac{m_{D_{s}}^{2}f_{D_{s}}}{m_{c}+m_{s}%
},\   \notag \\
\langle 0|J^{D_{s0}}|D_{s0}^{\ast }\rangle &=&m_{D_{s0}}f_{D_{s0}},
\label{eq:Mel2}
\end{eqnarray}%
where $f_{D_{s}}$ and $f_{D_{s0}}$ are the decay constants of the mesons $%
D_{s}^{+}$ and $D_{s0}^{\ast +}(2317)$, respectively. \ We also use the
following parametrization for the vertex
\begin{equation}
\langle D_{s}(q)D_{s0}^{\ast }(p^{\prime })|T(p)\rangle =G_{s}p\cdot
p^{\prime }  \label{eq:Ver1}
\end{equation}%
After some calculations it is not difficult to show that
\begin{eqnarray}
\Pi ^{\mathrm{Phys}}(p,p^{\prime }) &=&G_{s}\frac{%
m_{D_{s0}}f_{D_{s0}}m_{D_{s}}^{2}f_{D_{s}}m_{T}^{2}f_{T}}{\mathcal{M}%
^{2}(p^{\prime 2}-m_{D_{s0}}^{2})(p^{2}-m_{T}^{2})(q^{2}-m_{D_{s}}^{2})}
\notag \\
&&\times \left( m_{T}^{2}+m_{D_{s0}}^{2}-q^{2}\right) +\ldots .
\label{eq:Phys3}
\end{eqnarray}%
Because the Lorentz structure of the $\Pi ^{\mathrm{Phys}}(p,p^{\prime })$
is proportional to $I$, the invariant amplitude $\Pi ^{\mathrm{Phys}%
}(p^{2},p^{\prime 2},q^{2})$ is given exactly by Eq.\ (\ref{eq:Phys3}). Its
double Borel transformation over the variables $p^{2}$ and $p^{\prime 2}$
with the parameters $M_{1}^{2}$ and $M_{2}^{2}$ constitutes the left side of
the sum rule equality. Its right hand side is determined by the Borel
transformation $\mathcal{B}\Pi ^{\mathrm{OPE}}(p^{2},p^{\prime 2},q^{2})$,
where $\Pi ^{\mathrm{OPE}}(p^{2},p^{\prime 2},q^{2})$ is the invariant
amplitude that corresponds to the structure $\sim I$ in $\Pi ^{\mathrm{OPE}%
}(p,p^{\prime })$. Explicit expression of the correlation function $\Pi ^{%
\mathrm{OPE}}(p,p^{\prime })$ in terms of the quark propagators is presented
in the Appendix.

Equating $\mathcal{B}\Pi ^{\mathrm{OPE}}(p^{2},p^{\prime 2},q^{2})$ with the
double Borel transformation of $\Pi ^{\mathrm{Phys}}(p^{2},p^{\prime
2},q^{2})$ and performing continuum subtraction we get sum rule for the
strong coupling $G_{s}$, which is a function of $q^{2}$ and depends also on
the auxiliary parameters of calculations
\begin{eqnarray}
&&G_{s}(M^{2},\ s_{0},~q^{2})=\frac{\mathcal{M}^{2}}{%
m_{D_{s0}}f_{D_{s0}}m_{D_{s}}^{2}f_{D_{s}}m_{T}^{2}f_{T}}  \notag \\
&&\times \frac{q^{2}-m_{D_{s}}^{2}}{\left(
m_{T}^{2}+m_{D_{s0}}^{2}-q^{2}\right) }\int_{\mathcal{M}^{2}}^{s_{0}}ds\int_{%
\widetilde{\mathcal{M}}^{2}}^{s_{0}^{\prime }}ds^{\prime }\rho
_{s}(s,s^{\prime },q^{2})  \notag \\
&&\times e^{(m_{T}^{2}-s)/M_{1}^{2}}e^{(m_{D_{s0}}^{2}-s^{\prime
})/M_{2}^{2}},  \label{eq:SCoupl}
\end{eqnarray}%
where $\widetilde{\mathcal{M}}^{2}=\mathcal{M}^{2}/4$, and $\mathbf{M}%
^{2}=(M_{1}^{2},\ M_{2}^{2})$ and $\mathbf{s}_{0}=(s_{0},\ s_{0}^{\prime })$
are the Borel and continuum thresholds parameters, respectively.

One can see that the sum rule (\ref{eq:SCoupl}) is presented in terms of the
spectral density $\rho _{s}(s,s^{\prime },q^{2})$ which is proportional to
the imaginary part of $\Pi ^{\mathrm{OPE}}(p,p^{\prime })$. We calculate the
correlation function $\Pi ^{\mathrm{OPE}}(p,p^{\prime })$ by including
nonperturbative terms up to dimension six. But after double Borel
transformation only $s$-quark and gluon vacuum condensates $\langle \bar{s}%
s\rangle $ and $\langle \alpha _{s}G^{2}/\pi \rangle $ contribute to
spectral density $\rho _{s}(s,s^{\prime },q^{2})$, where, nevertheless, the
perturbative component plays a dominant role.

The strong coupling form factor $G_{s}(M^{2},\ s_{0},~q^{2})$ can be
calculated using the sum rule given by Eq.\ (\ref{eq:SCoupl}). The values of
the masses and decay constants of the mesons that enter into this expression
are collected in Table \ref{tab:Param}. Requirements which should be
satisfied by the auxiliary parameters $\mathbf{M}^{2}$ and $\mathbf{s}_{0}$
are similar to ones discussed in the previous section and are universal for
all sum rules computations. Performed analysis demonstrates that the working
regions
\begin{eqnarray}
M_{1}^{2} &=&(5,\ 7)\ \mathrm{GeV}^{2},\ s_{0}=(22,\ 24)\ \mathrm{GeV}^{2},
\notag \\
M_{2}^{2} &=&(3,\ 6)\ \mathrm{GeV}^{2},\ s_{0}^{\prime }=(7,\ 9)\ \mathrm{GeV%
}^{2},  \label{eq:Reg3}
\end{eqnarray}%
lead to stable results for the form factor $G_{s}(M^{2},\ s_{0},~q^{2})$,
and therefore are appropriate for our purposes. In what follows we omit its
dependence on the parameters and introduce $q^{2}=-Q^{2}$ denoting the
obtained form factor as $G_{s}(Q^{2})$.

\begin{figure}[h]
\includegraphics[width=8.8cm]{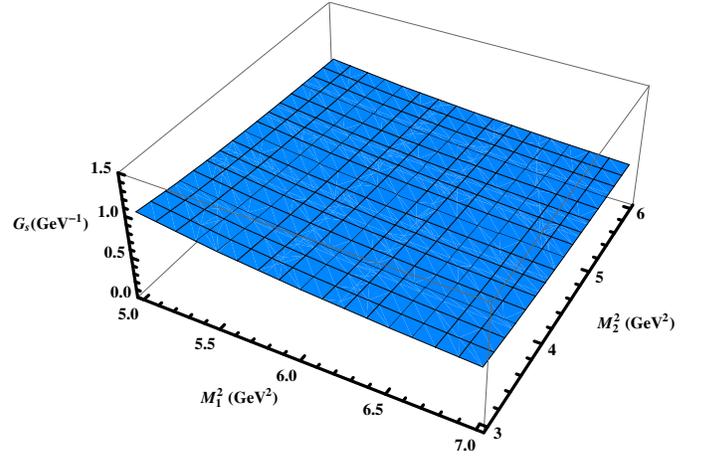}
\caption{The strong coupling form factor $G_s(Q^2)$ as a function of the
Borel parameters $\mathbf{M}^{2}=(M_{1}^2,\ M_{2}^2 )$ at the middle point
of the region $\mathbf{s}_0=(s_0, s_0^{\prime})$ and at fixed $Q^2=4~\mathrm{%
GeV}^2$.}
\label{fig:3Dplot}
\end{figure}

In order to visualize a stability of the sum rule calculations we depict in
Fig.\ \ref{fig:3Dplot} the strong coupling $G_{s}(Q^{2})$ as a function of
the Borel parameters at fixed $s_{0}$ and $Q^{2}$. It is seen that there is
a weak dependence of $G_{s}(Q^{2})$ on $M_{1}^{2}$ and $M_{2}^{2}$. The
dependence of $G_{s}(Q^{2})$ on $\mathbf{M}^{2}$, and also its variations
caused by the continuum threshold parameters are main sources of ambiguities
in sum rule calculations, which should not exceed $30\%$.

\begin{figure}[h]
\includegraphics[width=8.5cm]{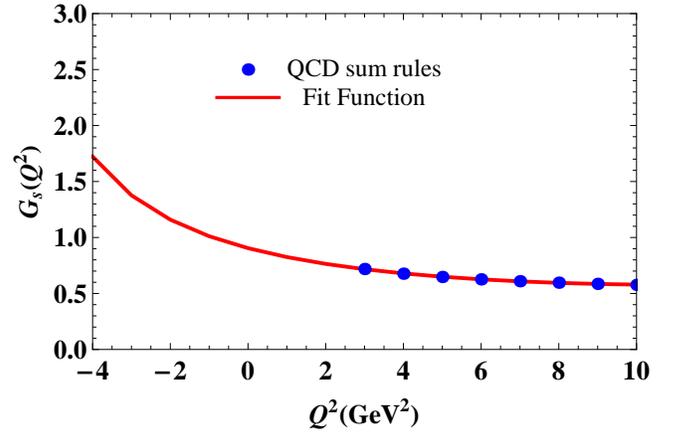}
\caption{The sum rule prediction and fit function for the strong coupling $%
G_{s}(Q^{2})$.}
\label{fig:Fit}
\end{figure}

For calculation of the decay width we need a value of the strong coupling at
the $D_{s}$ meson's mass shell, i.e. at $q^{2}=m_{D_{s}}^{2}$ or at $%
Q^{2}=-m_{D_{s}}^{2}$, where the sum rule method is not applicable.
Therefore it is necessary to introduce a fit function $F(Q^{2})$ that for
the momenta $Q^{2}>0$ leads to the same results as the sum rule, but can be
easily extended to the region of $Q^{2}<0$. It is convenient to model it in
the form
\begin{equation}
F(Q^{2})=\frac{f_{0}}{1-a\left( Q^{2}/m_{T}^{2}\right) +b\left(
Q^{2}/m_{T}^{2}\right) ^{2}},  \label{eq:Fitfunction}
\end{equation}%
where $f_{0}$, $a$ and $b$ are fitting parameters. The performed analysis
allows us to fix these parameters as $f_{0}=0.91\ \mathrm{GeV}^{-1}$, $%
a=-1.94$ and $b=-1.65$. The fit function $F(Q^{2})$ and sum rule results for
$G_{s}(Q^{2})$ are plotted in Fig.\ \ref{fig:Fit}, where one can see a very
nice agreement between them.

At the mass shell $Q^{2}=-m_{D_{s}}^{2}$ the strong coupling is equal to
\begin{equation}
G_{s}(-m_{D_{s}}^{2})=(1.67\pm 0.43)\ \mathrm{GeV}^{-1}.  \label{eq:Coupl1}
\end{equation}%
The width of the decay $T_{cc;\overline{s}\overline{s}}^{++}\rightarrow
D_{s}^{+}D_{s0}^{\ast +}(2317)$ is determined by the following formula
\begin{equation}
\Gamma \lbrack T_{cc;\overline{s}\overline{s}}^{++}\rightarrow
D_{s}^{+}D_{s0}^{\ast +}(2317)]=\frac{G_{s}^{2}m_{D_{s0}}^{2}}{8\pi }\lambda
\left( 1+\frac{\lambda ^{2}}{m_{D_{s0}}^{2}}\right) ,  \label{eq:DW}
\end{equation}%
where
\begin{eqnarray}
\lambda &=&\lambda \left( m_{T}^{2},m_{D_{s0}}^{2},m_{D_{s}}^{2}\right) =%
\frac{1}{2m_{T}}\left[ m_{T}^{4}+m_{D_{s0}}^{4}+m_{D_{s}}^{4}\right.  \notag
\\
&&\left.
-2(m_{T}^{2}m_{D_{s0}}^{2}+m_{T}^{2}m_{D_{s}}^{2}+m_{D_{s0}}^{2}m_{D_{s}}^{2})
\right] ^{1/2}.  \label{eq:Lambda}
\end{eqnarray}%
Our result for the decay width is:%
\begin{equation}
\Gamma =(302\pm 113)~\mathrm{MeV.}  \label{eq:Width1}
\end{equation}

In the similar calculations of the strong coupling $G_{d}(Q^{2})$ for the
Borel and threshold parameters $M_{1}^{2}$ and $s_{0}$ we have employed
\begin{equation}
M_{1}^{2}=(4.7,\ 6.5)\ \mathrm{GeV}^{2},\ s_{0}=(21,\ 23)\ \mathrm{GeV}^{2},
\label{eq:Reg4}
\end{equation}%
whereas $M_{2}^{2}$ and $\ s_{0}^{\prime }$ have been chosen as in Eq.\ (\ref%
{eq:Reg3}). For the strong coupling we have got
\begin{equation}
|G_{d}(-m_{D}^{2})|=(1.37\pm 0.34)\ \mathrm{GeV}^{-1}.  \label{eq:Coupl2}
\end{equation}%
Then the width of the process $T_{cc;\overline{d}\overline{s}%
}^{++}\rightarrow D^{+}D_{s0}^{\ast +}(2317)$ is
\begin{equation}
\widetilde{\Gamma }=(171~\pm 52)~\mathrm{MeV}.  \label{eq:Width2}
\end{equation}%
The predictions for the widths $\Gamma $ and $\widetilde{\Gamma }$ are the
final results of this section.
\begin{table}[tbp]
\begin{tabular}{|c|c|}
\hline\hline
Parameters & Values (in $\mathrm{MeV}$ units) \\ \hline\hline
$m_{D}$ & $1869.65\pm 0.05$ \\
$f_{D}$ & $211.9 \pm 1.1$ \\
$m_{D_s}$ & $1968.34\pm 0.07$ \\
$f_{D_s}$ & $249.0 \pm 1.2 $ \\
$m_{D_{s0}}$ & $2317.7\pm 0.6$ \\
$f_{D_{s0}}$ & $201$ \\ \hline\hline
\end{tabular}%
\caption{Parameters of the $D$-mesons used in numerical computations.}
\label{tab:Param}
\end{table}


%%%%%%%%%%%%%%%%%%%%%%%%%%%%%%%%%%%%%%%%%%%%%%%%%%%%%%%%%%%%%%%%%

\section{Analysis and concluding remarks}

\label{sec:Conc}
%%%%%%%%%%%%%%%%%%%%%%%%%%%%%%%%%%%%%%%%%%%%%%%%%%%%%%%%%%%%%%%%%%
In the present work we have calculated the spectroscopic parameters of the
doubly charmed $J^P=0^{-}$ tetraquarks $T_{cc;\overline{s}\overline{s}}^{++}$
and $T_{cc;\overline{d}\overline{s}}^{++}$ using QCD two-point sum rule
approach. Obtained results for mass of these resonances $m_{T}=(4390~\pm
150)~\mathrm{MeV}$ and $\widetilde{m}_{T}=(4265~\pm 140)~\mathrm{MeV}$
demonstrate that they are unstable particles and lie above open charm
thresholds $D_{s}^{+}D_{s0}^{\ast +}(2317)$ and $D^{+}D_{s0}^{\ast +}(2317)$%
, respectively. The mass splitting between the tetraquarks
\begin{equation}
m_{T}-\widetilde{m}_{T}\sim 125\ ~\mathrm{MeV}
\end{equation}%
is equal approximately to a half of mass difference between the ground-state
particles from $[cs][\overline{c}\overline{s}]$ and $[cq][\overline{c}%
\overline{q}]$ or from $[cs][\overline{b}\overline{s}]$ and $[cq][\overline{b%
}\overline{q}]$ multiplets \ \cite{Agaev:2017uky}. The quark content of
these resonances differs from each other by a pair of quarks $s\overline{s}$
and $q\overline{q}$, whereas the tetraquark $T_{cc;\bar{d}\bar{s}}^{++}$ can
be obtained from $T_{cc;\bar{s}\bar{s}}^{++}$ by only $\overline{s}%
\rightarrow \overline{d}$ replacement. In other words, the mass splitting
caused by the $s$-quark equals to $125\ ~\mathrm{MeV}$. It is interesting
that in the conventional mesons $s$-quark's "mass" is lower and amounts to $%
D_{s}^{+}(c\overline{s})-D^{0}(c\overline{u})\approx 100\ ~\mathrm{MeV}$ ,
whereas for baryons, for example $\Xi _{c}^{+}(usc)-\Lambda
_{c}^{+}(udc)\approx 180\ ~\mathrm{MeV}$, it is higher than $125\ ~\mathrm{%
MeV}$.

We have also evaluated the widths of the tetraquarks $T_{cc;\overline{s}%
\overline{s}}^{++}$ and $T_{cc;\overline{d}\overline{s}}^{++}$ through their
dominant $S$-wave strong decays to the pair of $D_{s}^{+}D_{s0}^{\ast
+}(2317)$ and $D^{+}D_{s0}^{\ast +}(2317)$ mesons. To this end we have
employed QCD three-point sum rules approach and found the strong couplings $%
G_{s}$ and $G_{d}:$ they are key ingredients of computations. The widths $%
\Gamma =(302~\pm 113)~\mathrm{MeV}$ and $\widetilde{\Gamma }=(171~\pm 52)~%
\mathrm{MeV}$ show that the tetraquarks $T_{cc;\overline{s}\overline{s}%
}^{++} $ and $T_{cc;\overline{d}\overline{s}}^{++}$ can be classified as
rather broad resonances.

We have evaluated the spectroscopic parameters of the tetraquarks $T_{cc;%
\overline{s}\overline{s}}^{++}$ and $T_{cc;\overline{d}\overline{s}}^{++}$
using the zero-width single-pole approximation. But, as it has been
emphasized in the section \ref{sec:Mass}, the interpolating currents (\ref%
{eq:Curr1}) and (\ref{eq:Curr2}) couple not only to the tetraquarks, but
also to the two-meson continuum (in our case, to the states $%
D_{s}^{+}D_{s0}^{\ast +}(2317)$ and $D^{+}D_{s0}^{\ast +}(2317)$), and these
effects may correct our predictions for $m_{T}$, $f_{T}$ and $\widetilde{m}%
_{T}$, $\widetilde{f}_{T}$, respectively. The two-meson continuum
contribution modifies the zero-width approximation (\ref{eq:Mod}) and in the
case of the tetraquark $T_{cc;\overline{s}\overline{s}}^{++}$ leads to the
following corrections \cite{Wang:2015nwa}:%
\begin{equation}
\lambda _{T}^{2}e^{-m_{T}^{2}/M^{2}}\rightarrow \lambda
_{T}^{2}\int_{(m_{D_{s}}+m_{D_{s0}})^{2}}^{s_{0}}dsW(s)e^{-s/M^{2}}
\label{eq:Correction1}
\end{equation}%
and
\begin{equation}
\lambda _{T}^{2}m_{T}^{2}e^{-m_{T}^{2}/M^{2}}\rightarrow \lambda
_{T}^{2}\int_{(m_{D_{s}}+m_{D_{s0}})^{2}}^{s_{0}}dsW(s)se^{-s/M^{2}},
\label{eq:Correction2}
\end{equation}%
where $\lambda _{T}^{2}=m_{T}^{4}f_{T}^{2}/\mathcal{M}^{2}.$ In Eqs.\ (\ref%
{eq:Correction1}) and (\ref{eq:Correction2}) we have used
\begin{equation}
W(s)=\frac{1}{\pi }\frac{m_{T}\Gamma (s)}{\left( s-m_{T}^{2}\right)
^{2}+m_{T}^{2}\Gamma ^{2}(s)}  \label{eq:NewF1}
\end{equation}%
and
\begin{equation}
\Gamma (s)=\Gamma \frac{m_{T}}{s}\sqrt{\frac{s-(m_{D_{s}}+m_{D_{s0}})^{2}}{%
m_{T}^{2}-(m_{D_{s}}+m_{D_{s0}})^{2}}}.  \label{eq:NewF2}
\end{equation}%
By utilizing the central values of the $m_{T}$ and $\Gamma $, as well as $%
M^{2}=6\ \mathrm{GeV}^{2}$ and $s_{0}=23\ \mathrm{GeV}^{2}$ it is not
difficult to find that%
\begin{equation}
\lambda _{T}^{2}\rightarrow 0.86\lambda _{T}^{2}\rightarrow \frac{%
(0.927f_{T})^{2}m_{T}^{4}}{\mathcal{M}^{2}},  \label{eq:Rescale1}
\end{equation}%
and
\begin{equation}
\lambda _{T}^{2}m_{T}^{2}\rightarrow 0.86\lambda
_{T}^{2}m_{T}^{2}\rightarrow \frac{(0.927f_{T})^{2}m_{T}^{6}}{\mathcal{M}^{2}%
}.  \label{eq:Rescale2}
\end{equation}%
As is seen, in both cases the two-meson effects result in rescaling of the
coupling $f_{T}\rightarrow 0.927f_{T}$, i.e., change it approximately by $%
7.3\%$ and do not exceeds the accuracy of the sum rule calculations which
amounts to $\pm 19\%$. The similar estimation $\widetilde{f}_{T}\rightarrow
0.945\widetilde{f}_{T}$ is valid for the coupling $\widetilde{f}_{T}$ as
well.

The double-charmed tetraquarks investigated in the present work carry a
double electric charge and may exist as diquark-antidiquarks. They are
unstable resonances, but some of double-bottom tetraquarks may be stable
against strong decays. Therefore theoretical and experimental studies of the
double-heavy four-quark systems, their strong and weak decays remain in the
agenda of high energy physics, and can provide valuable information on
internal structure and properties of these exotic mesons.

%%%%%%%%%%%%%%%%%%%%%%%%%%%%%%%%%%%%%%%%%%%%%%%%%%%%%%%%%%%%%%%%%%%%%%%%%%

\section*{ACKNOWLEDGEMENTS}

%%%%%%%%%%%%%%%%%%%%%%%%%%%%%%%%%%%%%%%%%%%%%%%%%%%%%%%%%%%%%%%%%%%%%%%%%
K.~A., B.~B.~ and H.~S.~ thank TUBITAK for the financial support provided
under Grant No. 115F183.

%%%%%%%%%%%%%%%%%%%%%%%%%%%%%%%%%%%%%%%%%%%%%%%%%%%%%%%%%%%%%%%%%%%%%%%%
\appendix*

\section{ The correlation functions used in calculations}

\renewcommand{\theequation}{\Alph{section}.\arabic{equation}} \label{sec:App}
%%%%%%%%%%%%%%%%%%%%%%%%%%%%%%%%%%%%%%%%%%%%%%%%%%%%%%%%%%%%%%%%%%%%%%

In this Appendix we have collected the explicit expressions of the
correlation functions $\Pi ^{\mathrm{OPE}}(p)$ and $\Pi ^{\mathrm{OPE}%
}(p,p^{\prime })$ used in the sections \ref{sec:Mass} and \ref{sec:Decays}
to derive sum rules for calculation of the spectroscopic parameters of the
tetraquark $T_{cc;\overline{s}\overline{s}}^{++}$ and its decay width. The
function $\Pi ^{\mathrm{OPE}}(p)$ has the following expression in terms of
the quark propagators:
\begin{widetext}

\begin{eqnarray}
&&\Pi ^{\mathrm{OPE}}(p)=-2i\int d^{4}xe^{ipx}\left\{ \mathrm{Tr}\left[
S_{c}^{bb^{\prime }}(x)\widetilde{S}_{c}^{aa^{\prime }}(x)\right] \mathrm{Tr}%
\left[ \gamma _{5}\widetilde{S}_{s}^{a^{\prime }b}(-x)\gamma
_{5}S_{s}^{b^{\prime }a}(-x)\right] +\mathrm{Tr}\left[ S_{c}^{ba^{\prime
}}(x)\widetilde{S}_{c}^{ab^{\prime }}(x)\right] \right.  \notag \\
&&\times \mathrm{Tr}\left[ \gamma _{5}\widetilde{S}_{s}^{a^{\prime
}b}(-x)\gamma _{5}S_{s}^{b^{\prime }a}(-x)\right] +\mathrm{Tr}\left[
S_{c}^{bb^{\prime }}(x)\widetilde{S}_{c}^{aa^{\prime }}(x)\right] \mathrm{Tr}%
\left[ \gamma _{5}\widetilde{S}_{s}^{b^{\prime }b}(-x)\gamma
_{5}S_{s}^{a^{\prime }a}(-x)\right]  \notag \\
&&+\mathrm{Tr}\left[ S_{c}^{ba^{\prime }}(x)\widetilde{S}_{c}^{ab^{\prime
}}(x)\right] \mathrm{Tr}\left[ \gamma _{5}\widetilde{S}_{s}^{b^{\prime
}b}(-x)\gamma _{5}S_{s}^{a^{\prime }a}(-x)\right] +\mathrm{Tr}\left[
S_{c}^{bb^{\prime }}(x)\widetilde{S}_{c}^{aa^{\prime }}(x)\right] \mathrm{Tr}%
\left[ \gamma _{5}\widetilde{S}_{s}^{a^{\prime }a}(-x)\gamma
_{5}S_{s}^{b^{\prime }b}(-x)\right]  \notag \\
&&+\mathrm{Tr}\left[ S_{c}^{ba^{\prime }}(x)\widetilde{S}_{c}^{ab^{\prime
}}(x)\right] \mathrm{Tr}\left[ \gamma _{5}\widetilde{S}_{s}^{a^{\prime
}a}(-x)\gamma _{5}S_{s}^{b^{\prime }b}(-x)\right] +\mathrm{Tr}\left[
S_{c}^{bb^{\prime }}(x)\widetilde{S}_{c}^{aa^{\prime }}(x)\right] \mathrm{Tr}%
\left[ \gamma _{5}\widetilde{S}_{s}^{b^{\prime }a}(-x)\gamma
_{5}S_{s}^{a^{\prime }b}(-x)\right]  \notag \\
&&\left. +\mathrm{Tr}\left[ S_{c}^{ba^{\prime }}(x)\widetilde{S}%
_{c}^{ab^{\prime }}(x)\right] \mathrm{Tr}\left[ \gamma _{5}\widetilde{S}%
_{s}^{b^{\prime }a}(-x)\gamma _{5}S_{s}^{a^{\prime }b}(-x)\right] \right\} ,
\end{eqnarray}%
where
\begin{equation*}
\widetilde{S}_{c(s)}(x)=CS_{c(s)}^{\mathrm{T}}(x)C.
\end{equation*}%
Here $S_{c(s)}(x)$ is the heavy $c$-quark ( light $s$-quark) propagator.

The correlation function $\Pi ^{\mathrm{OPE}}(p,p^{\prime })$ is presented
below
\begin{eqnarray}
&&\Pi ^{\mathrm{OPE}}(p,p^{\prime })=2i^{2}\int
d^{4}xd^{4}ye^{-ipx}e^{ip^{\prime }y}\left\{ \mathrm{Tr}\left[
S_{c}^{jb^{\prime }}(y-x)\widetilde{S}_{c}^{ia^{\prime }}(-x)\gamma _{5}%
\widetilde{S}_{s}^{b^{\prime }i}(x)\gamma _{5}S_{s}^{a^{\prime }j}(x-y)%
\right] \right.  \notag \\
&&+\mathrm{Tr}\left[ S_{c}^{ja^{\prime }}(y-x)\widetilde{S}_{c}^{ib^{\prime
}}(-x)\gamma _{5}\widetilde{S}_{s}^{b^{\prime }i}(x)\gamma
_{5}S_{s}^{a^{\prime }j}(x-y)\right] +\mathrm{Tr}\left[ S_{c}^{jb^{\prime
}}(y-x)\widetilde{S}_{c}^{ia^{\prime }}(-x)\gamma _{5}\right.  \notag \\
&&\left. \left. \times \widetilde{S}_{s}^{a^{\prime }i}(x)\gamma
_{5}S_{s}^{b^{\prime }j}(x-y)\right] +\mathrm{Tr}\left[ S_{c}^{ja^{\prime
}}(y-x)\widetilde{S}_{c}^{ib^{\prime }}(-x)\gamma _{5}\widetilde{S}%
_{s}^{a^{\prime }i}(x)\gamma _{5}S_{s}^{b^{\prime }j}(x-y)\right] \right\} .
\end{eqnarray}

\end{widetext}

\end{document}